\documentclass[twocolumn,showpacs,amsmath,amssymb,pra,superscriptaddress]{revtex4}
\usepackage{epsfig}
\usepackage{color}
\usepackage{graphicx}
\usepackage{bm}
\usepackage{epstopdf}

\begin{document}

\title{Quantum magnetism in strongly interacting one-dimensional spinor Bose systems}

\author{A.~S. Dehkharghani}
\author{A.~G. Volosniev}
\affiliation{Department of Physics and Astronomy, Aarhus University,
  DK-8000 Aarhus C, Denmark} 
\author{E.~J. Lindgren}
\author{J. Rotureau}
\affiliation{Department of Fundamental Physics, Chalmers University of
  Technology, SE-412 96 G{\"o}teborg, Sweden} 
\author{C. Forss{\'e}n}
\affiliation{Department of Fundamental Physics, Chalmers University of
  Technology, SE-412 96 G{\"o}teborg, Sweden} 
\affiliation{Department of Physics and Astronomy, University of Tennessee, Knoxville, TN 37996, USA}
\affiliation{Physics Division, Oak Ridge National Laboratory, Oak Ridge, TN 37831, USA}
\author{D.~V. Fedorov} 
\author{A.~S. Jensen}
\author{N.~T. Zinner}
\affiliation{Department of Physics and Astronomy, Aarhus University, 
DK-8000 Aarhus C, Denmark}

\date{\today}

\begin{abstract}
Strongly interacting one-dimensional quantum systems often behave in a 
manner that is distinctly different from their higher-dimensional counterparts.
When a particle attempts to move in a one-dimensional environment it will 
unavoidably have to interact and 'push' other particles in order to execute
a pattern of motion, irrespective of whether the particles are fermions or 
bosons. A present frontier in both theory and experiment are mixed systems 
of different species and/or particles with multiple internal degrees of 
freedom.
Here we consider trapped two-component bosons with short-range inter-species interactions much larger 
than their intra-species interactions and show that they have
novel energetic and magnetic properties. 
In the strongly interacting regime, these systems have 
energies that are fractions of the basic harmonic oscillator trap quantum
and have spatially separated ground states with manifestly ferromagnetic 
wave functions. Furthermore, we predict excited states that have perfect
antiferromagnetic ordering. This holds for both balanced and imbalanced 
systems, and we show that it is a generic feature as one crosses from 
few- to many-body systems.
\end{abstract}


\maketitle

\section{Introduction}
The interest in one-dimensional (1D) quantum systems with several interacting
particles arguably began back in 1931 when Bethe solved the famous 
Heisenberg model of ferromagnetism \cite{bethe1931}, but it was only
in the 1960s that people realized that the techniques invented by
Bethe could be used to solve a host of different many-body models 
\cite{lieb1963,mcguire1965,yang1967,lieb1968}. It was subsequently 
realized that many 1D systems have universal low-energy behaviour 
and can be described by the paradigmatic Tomonaga-Luttinger-Liquid (TLL)
theory \cite{haldane1981a,haldane1981b,giamarchi2004}.
This opened up the 
field of one-dimensional physics, which has remained a large subfield
of condensed-matter physics ever since \cite{giamarchi2004,cazalilla2011}. 
Recently, there has been a great revival of interest in 1D systems
due to the realization of 1D quantum gases in highly controllable
environments using cold atomic gases \cite{paredes2004,kino2004,kinoshita2006,haller2009,serwane2011,gerhard2012,wenz2013}.
This development implies that one may now experimentally realize 1D systems
with bosons or fermions and explore the intricate nature of their 
quantum behaviour.

A recent frontier is the realization of multi-component systems \cite{pagano2014}
in order to study fundamental 1D effects such as spin-charge 
separation \cite{recati2003}. While this effect is usually 
associated with spin 1/2 fermions, it turns out that
it can also be explored in Bose mixtures (two-component bosonic systems)
where the phenomenon can be even richer as there can be interactions 
between the two components (inter-species) and also within each 
component separately (intra-species) \cite{kuklov2003,duan2003,cazalilla2011}.
The latter is strongly suppressed for fermions due to the Pauli principle.
In the case where the intra- and inter-species interactions are
identical it has been shown that a ferromagnetic ground state
occurs \cite{eisenberg2002,nachtergaele2005}.
Generalizing to the case of unequal intra-
and inter-species interactions may be possible, but since the
proofs and techniques rely on spin algebra
and representation theory, they cannot be used to obtain the
full spatial structure of general systems
and other approaches are therefore needed.
Here we consider the limit where the inter-species 
dominates the intra-species interactions. This regime has been 
explored in recent years for small systems using various few-body techniques 
\cite{zollner2008,hao2009,garciamarch2013a,garciamarch2013b,garciamarch2013c,zinner2013}
and behaviour different from strongly interacting fermions or 
single-component bosons can be found already for
three particles \cite{zinner2013}. 
From the many-body side, 
the system is known to have spin excitations with quadratic
dispersion, \cite{sutherland1968,li2003,fuchs2005,guan2007}
which can be shown to be a generic feature of the 'magnon' 
excitations above a ferromagnetic ground state \cite{halperin1969,halperin1975}.
This goes beyond the TLL theory and it has been conjectured
that a new universality class ('ferromagnetic liquid') emerges in 
this regime \cite{zvonarev2007,akhanjee2007,matveev2008,kamenev2009,caux2009}.

Here we provide a particularly clean realization of a 'ferromagnetic' system confined
in a harmonic trap. Using numerical and newly developed analytical 
techniques we obtain and analyze the exact 
wave function. 
This allows us to explore the crossover between few- and 
many-body behaviour, and to demonstrate that the strongly interacting
regime realizes a perfect ferromagnet in the ground state, while 
particular excited states will produce perfect antiferromagnetic order.
In the extremely imbalanced system, with one strongly interacting 
'impurity', we find both numerically and analytically that the 
impurity will always move to the edge of the system. This is 
in sharp contrast to fermionic systems where the impurity is mainly 
located at the center \cite{lindgren2014}.
Our work provides a rare and explicit example of perfect ferro- or 
antiferromagnetism using the most fundamental knowledge of a
quantum system as given by the full wave function.

\begin{figure}[ht!]
\centering
\includegraphics[scale=0.45,clip=true]{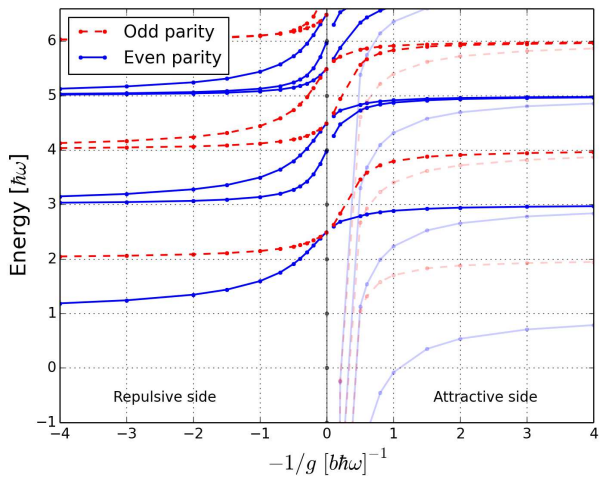}
\caption{Three-body spectral flow. The energy spectrum of two $A$ and one $B$ particle as a function of 
interaction strength, $g$, obtained by numerical calculations. In the limit $1/g\to 0$, the ground state
becomes doubly degenerate and has half-integer energy. 
The contribution from center-of-mass motion has been removed. For visibility, we have removed states from the 
attractive side that diverge to large negative energies close to $1/g\to 0$.}
\label{2+1}
\end{figure}

\section{Energetics and wave functions} 
Our two-component bosonic system has
$N=N_A+N_B$ particles split between $N_A$ and $N_B$ identical bosons of 
two different kinds. All $N$ particles have mass $m$ and move in the same
external harmonic trapping potential with single-particle Hamiltonian
$h_0=\frac{p^2}{2m}+\frac{1}{2}m\omega^2 x^2$,
where $p$ and $x$ denote the momentum and position of either an $A$ or $B$ particle and 
$\omega$ is the common trap frequency. The trap provides a natural set
of units for length, $b=\sqrt{\hbar/m\omega}$, and energy, $\hbar\omega$, which
we will use throughout (here $\hbar$ is Planck's constant divided by $2\pi$).
We assume short-range interactions
between $A$ and $B$ particles that we model by a Dirac delta-function 
parameterized by an interaction strength, $g$, i.e. 
\begin{align}\label{interaction}
H_I=g\sum_{i=1}^{N_A}\sum_{j=1}^{N_B}\delta(x_{i}-y_{j}),
\end{align}
where $x$ and $y$ denote the coordinates of $A$ and $B$ particles, respectively.
The intraspecies interaction strengths are assumed to be much smaller than $g$
and we will therefore neglect such terms. To access the quantum mechanical 
properties of our system we must solve the $N$-body Schr{\"o}dinger equation. 
This will be done using novel analytical tools
and using exact diagonalization. In the latter case we have adapted an effective
interaction approach that has recently been succesfully applied to fermions
in harmonic traps \cite{rotureau2013,lindgren2014} (see the Methods section for further details). 
The analytical and numerical methods allow us to address up to ten particles,
which is larger than most previous studies not based on stochastic 
or Monte Carlo techniques.

The simplest non-trivial case is the three-body system which has 
two $A$ and one $B$ particle. The energy spectrum is shown in Fig.~\ref{2+1}
as a function of $g$. 
The most interesting feature to notice is the ground state behaviour as 
$1/g\to 0^+$. Here, an odd and an even state become degenerate at 
an energy of $2.5\hbar\omega$. This should be contrasted to the 
behaviour of single-component bosons or two-component fermions which
will always have energies that are an integer times $\hbar\omega$ when
$1/g\to 0$. Furthermore, we notice how the two states that merge
at $1/g=0$ become two excited state branches on the attractive side of 
the resonance but the even parity state remains the lower one. 
This is opposite to the behaviour of fermions \cite{volosniev2013} where the hierarchy 
of states is inverted at $1/g=0$.
The ground state for large and negative $g$ is very different
as it contains deeply bound molecules, which we will not consider
further.
The fractional energies in the spectrum can be 
explained by a schematic three-body model and 
in stochastic variational calculations \cite{zinner2013}.
This provides a hint that larger systems could also display fractional 
energy states in the strongly interacting limit and begs the question 
as to what spatial configurations such states correspond to. 

We 
will now show that the fractional energy states are generic
for strongly interacting two-component bosons in 1D and, importantly,
for the ground state they realize perfect ferromagnetic behaviour 
irrespective of whether the system is balanced ($N_A=N_B$) or not.
The term perfect ferromagnetic behaviour implies that we have a 
full spatial separation of the two components in the exact
ground state wave function of the system, i.e. the probability
to find only $A$ on one side and only $B$ on the other side of the
system is not just dominant, it is exactly {\it unity}. 
The 
ground state has only a single 'domain wall' at which an $A$ and a $B$
particle are neighbours. As a consequence, imagine that you detect 
an $A$ particle on the left (right) side of the system, then you 
can immediately conclude that all the $B$ particles reside
to the right (left) of this $A$ particle.

\begin{figure*}[ht!]
\centering
\includegraphics[scale=0.7,clip=true]{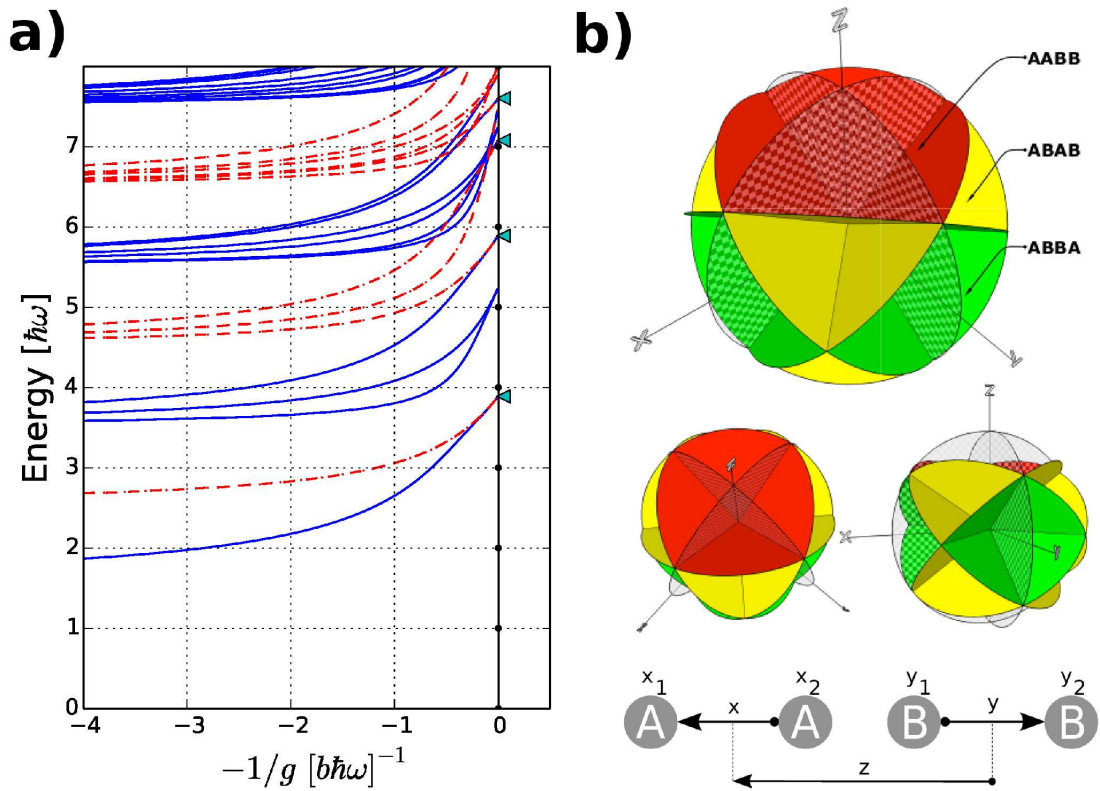}
\caption{Balanced four-body system. {\bf a}) Energy spectrum for $g>0$ for two $A$ and two $B$ particles. 
The $1/g\to 0^+$ limits are analytically known and indicated by triangles. Even parity states are in blue while odd
parity states are red. {\bf b}) Three-dimensional representation 
of the coordinate space on which the four-body wave function is defined when the center-of-mass position is
removed. The specific Jacobi coordinates used are shown at the bottom. 
The solid coloured circular planes indicate the planes where an $AB$ pair overlap.
The wave function must vanish on these planes in the limit where $1/g\to 0$. The checkerboard coloured 
circular planes are reflection planes for a pair of identical particles ($x\to -x$ and $y\to -y$). 
The red region has twice the 
volume of the green region and four times that of the yellow region. The smaller figures in the middle
show the same regions viewed from different angles for clarity.
}
\label{2+2}
\end{figure*}

\subsection{Balanced systems} 
We first consider a four-body system that has two $A$ 
and two $B$ particles. The energy spectrum for $g>0$ is shown in 
Fig.~\ref{2+2}{\bf a}). A striking feature is the two-fold 
degenerate ground state for $1/g\to 0$ that has a non-integer
energy similar to the three-body problem. In this strongly 
repulsive limit, the system realizes a perfect spatially ferromagnetic 
quantum state as we will now demonstrate analytically.  

First we note that the center-of-mass motion of the four-body system can 
be separated and thus ignored. This leaves three Jacobi coordinates to 
describe the system. The details of these reductions can be found 
in the Methods section below.
In Fig.~\ref{2+2}{\bf b}) we show the space
of the Jacobi coordinates and highlight all the planes 
at which an $AB$ (solid planes), an $AA$ or a $BB$ (checkerboard planes) pair of particles overlap.
The main observation is that as $1/g\to 0$, the wave function 
must vanish on all the solid planes in Fig.~\ref{2+2}{\bf b})
and we arrive at the disconnected regions shown with different colours. The particles 
become effectively impenetrable and we may characterize the wave function 
by specifying the amplitudes of all possible spatial configurations 
of the four particles. These regions 
correspond to specific orderings on a line of the four particles. 
In particular, the large (red) region dominating the figure corresponds
to spatial configurations $AABB$ or $BBAA$. The green region occupies 
half the spatial volume of the red and corresponds to $ABBA$ or $BAAB$, 
while the yellow region has one-fourth the volume of the red region and corresponds to 
$ABAB$ or $BABA$ configurations. A wave function that vanishes on all
$AB$ interfaces may now be constructed in each of these regions. However,
it is immediately clear that it will have lower energy when it can 
spread over a larger volume. We thus conclude that the 
doubly degenerate ground state at $1/g\to 0$ has the structure $AABB\pm BBAA$ (taking into 
account the parity invariance of the Hamiltonian). 

As discussed in the Methods section, one may solve a 
simple wave equation in the red region and obtain the 
ground state energy to arbitrary precision. 
The triangles at the $1/g=0$ line in Fig.~\ref{2+2}{\bf a})
show the energies obtained in this manner. We reproduce both the 
ground state and a set of excited states.
All of these have fractional energies and all 
of them are perfectly ferromagnetically ordered. The remaning states of the
spectrum can be obtained by solving in the other regions of Fig.~\ref{2+2}{\bf b}).
Note that states with amplitude exclusively in the yellow regions are 
perfectly spatial antiferromagnetic, $ABAB\pm BABA$, and have energies $n+1/2$ $\hbar\omega$
with integer $n$. 
They are the only parts of the spectrum which can be constructed 
by starting from identical fermions using Girardeau's mapping
techniques \cite{girardeau1960}. The arguments presented here are 
neither restricted to $N=4$ nor to a harmonic trapping potential and hold for 
any $N$ and any shape of the external confinement.
They hinge only on the fact
that the $AABB$ or $BBAA$ configurations occupy the largest
volume. For instance, for $AABB$ the four regions $x_1<x_2<y_1<y_2$,
$x_2<x_1<y_1<y_2$, $x_1<x_2<y_2<y_1$ and $x_2<x_1<y_2<y_1$ 
are adjacent regions and are connected by Bose symmetry of each component, 
thus they make up 
the connected upper red region ($z>0$) of the coordinate space 
in Fig.~\ref{2+2}{\bf b}).
For $ABBA$ one finds that only $x_1<y_1<y_2<x_2$ and $x_1<y_2<y_1<x_2$ are
adjacent and connected. This implies that the volume for $ABBA$
is half as large. 

\begin{figure}[ht!]
\centering
\includegraphics[scale=0.42,clip=true]{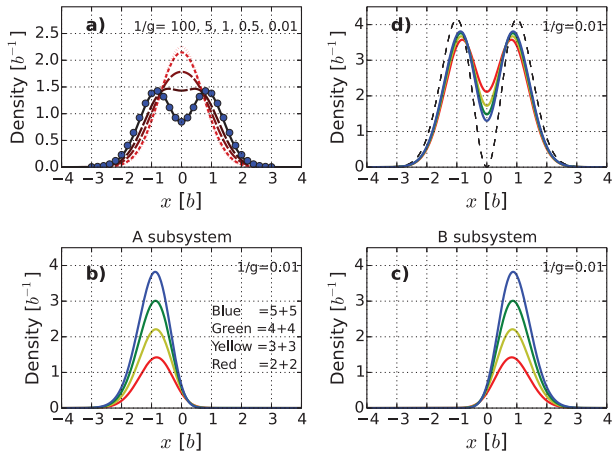}
\caption{Ground state densities of balanced systems. 
{\bf a}) Total density for $N_A=N_B=2$
and different values of $g$. Dotted (red) line corresponds to $1/g=100$ while the solid (black) line is for $1/g=0.01$.
The dots show the analytical solution for $1/g=0$. {\bf b}) and {\bf c}) Densities
for an equal superposition (sum) of the (nearly) two-fold degenerate ground state at $1/g=0.01$ for $N_A=N_B=2$, 3, 4, and 5.
{\bf b}) shows $A$ particles and {\bf c}) shows $B$ particles. {\bf d}) Rescaled plot of the total density at $1/g=0.01$. For $N_A=N_B<5$ the density has been rescaled to a total density 
with $N_A=N_B=5$. The dashed line corresponds to the density expected in the many-body limit $N_A=N_B\gg 1$.}
\label{bdens}
\end{figure}

Larger systems may in principle be handled in similar fashion by 
solving wave equations with proper boundary conditions and obtaining
the fractional energies in the limit $1/g\to 0$. However, the increase
in dimension of the problem makes this very difficult in practice. 
In order to further demonstrate that balanced systems have
perfect ferromagnetic ground states irrespective of particle 
number, we have numerically computed the ground state densities for systems
with $N_A=N_B\leq 5$ as shown in Fig.~\ref{bdens}. Evidence of the 
separation of $A$ and $B$ can be seen in the total density already
in Fig.~\ref{bdens}{\bf a}) for $N_A=N_B=2$ as $g$ increases
(note the perfect agreement with the analytical result in the 
limit $1/g\to 0$). 
We expect the two degenerate ground states to have structure 
$A\ldots AB\ldots B\pm B\ldots BA\ldots A$. In order to prove
this perfect ferromagnetic behaviour, we consider the odd and even
superposition of the two degenerate states which we expect will yield states
with exclusively $A$ or $B$ particles on either side of the system
(corresponding to $A\ldots AB\ldots B$ or $B\ldots BA\ldots A$).
The corresponding
densities are shown in Fig.~\ref{bdens}{\bf b}) and Fig.~\ref{bdens}{\bf c}) 
and beautifully confirm our expectations. 

As the ground state for $1/g\to 0$ is spatially separated, one may speculate that 
it can be understood physically as two ideal Bose gases or 'condensates' sitting on either
side of the system even in this strongly interacting limit. 
In Fig.~\ref{bdens}{\bf d}) we plot the densities
in a rescaled fashion where we multiply by $5/2$, $5/3$, and $5/4$ 
on the $N_A=N_B=2$, $3$ and $4$ densities respectively. The convergence
of the results toward the $N_A=N_B=5$ case indicates that the system does
behave as two ideal Bose gases as the particle number grows. 
In the limit $N_A=N_B\gg 1$, we would expect the overlap of the
two gases to vanish as the energy cost of overlap goes to infinity. 
We therefore expect that the occupied mode in this large
system limit is the first excited state of the harmonic trap which 
vanishes at the center. The dashed line in Fig.~\ref{bdens}{\bf d})
shows this state rescaled to $N_A=N_B=5$. This analytical guess
displays the same features as the numerical densities and we conclude
that already for ten particles the many-body properties are emerging.

\begin{figure}[ht!]
\centering
\includegraphics[scale=0.42,clip=true]{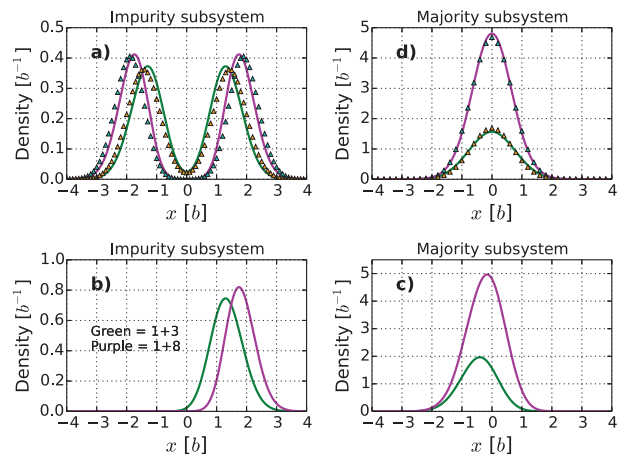}
\caption{Ground state densities of imbalanced systems. {\bf a}) Impurity density
in an $N_B=3$ or $N_B=8$ system with $N_A=1$. The analytical results for $1/g=0$ are shown as triangles.
{\bf b}) and {\bf c}) as in Fig.~\ref{bdens}{\bf b}) and {\bf c}) but for $N_A=1$ with $N_B=3$
or $N_B=8$. {\bf d}) The density of the majority component ($N_B$). All numerical 
results have been obtained with $1/g=0.01$.
}
\label{imdens}
\end{figure}

\subsection{Imbalanced systems} 
The extremely imbalanced limit, where $N_A=1$
and $N_B$ varies, provides a realization of a strongly interacting 
Bose polaron
in 1D, i.e. an impurity that interacts strongly with an ideal
Bose gas. In Fig.~\ref{imdens} we plot the densities of 
systems with $N_A=1$ and $N_B=3$ or $N_B=8$. We see that the 
impurity sits at the edge of the system (Fig.~\ref{imdens}{\bf a})),
while the majority component tends to occupy the center (Fig.~\ref{imdens}{\bf d})).
We confirm the numerical results by employing an analytical model, 
which shows excellent agreement. The details can be found in the 
Methods section. To confirm that the wave function of the strongly
interacting ground state has intrinsic phase separation, i.e. 
has the form $AB\ldots B\pm B\ldots BA$, we plot the densities for a sum
of the nearly degenerate ground states in Fig.~\ref{imdens}{\bf b}) 
and {\bf c}). As in the balanced case above, we find a perfectly
separated ground state behaviour. For instance, if we locate the single $A$ particle
on one side of the trap, we would thus immediately 
know that all the $B$ particles reside on the other side, and vice versa.
This behaviour is opposite to the case where the $B$ particles are 
identical fermions where
the impurity resides mainly in the center of the system \cite{lindgren2014}.
We have confirmed that this structure is also present for $N_A\neq N_B$ with $N_A>1$,
and it is therefore a generic feature that the two species are perfectly 
spatially separated (ferromagnetic) in the ground state for strong 
interactions.

\begin{figure}[ht!]
\centering
\includegraphics[scale=0.45,clip=true]{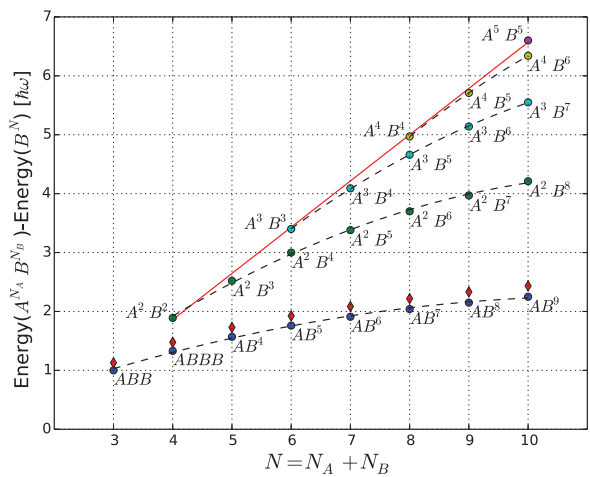}
\caption{Ground state energies for $N\leq 10$. The filled circles
show the ground state energy for $1/g=0.01$ relative to the zero-point energy 
as given by a single-component system of the same size. Each point is marked with 
the number of particles using the notation $A^{N_A} B^{N_B}$. The diamonds are the results
of the analytical method for the polaron case described in the Methods section. 
The dashed lines are quadratic interpolations for fixed number of $B$ particles, 
while the solid line is an interpolation of the energy for the balanced systems.
}
\label{enetot}
\end{figure}

A remarkable feature of the densities in Fig.~\ref{imdens}{\bf b})
and {\bf c}) is the movement of the centroids of the 
peaks with particle number. We clearly see the majority moving 
into the center and the impurity being pushed toward the edge. 
This demonstrates how an ideal condensate is being built in 
the center. For large $N_B$ the energy per particle goes to 
$1/2\hbar\omega$, which implies a single-mode condensate
that is becoming macroscopically occupied (see Methods section 
for details). The relative deviation between numerical and analytical 
energies is below three percent for $N_B\geq 9$. We thus have 
an analytic model for the crossover between the few- and 
many-body limit for the bosonic polaron in one dimension.
This includes the external trap that is a reality of most 
experiments. 

\section{Discussion}
We have shown that a mixture of two ideal 
Bose systems in one dimension has unusual properties 
when the inter-species interactions
is strong. The systems have 
energies that are non-integer multiples of $\hbar\omega$. In Fig.~\ref{enetot} we show the
ground state energies for $1/g=0.01$ in systems with ten particles 
or less relative to a ground state with only $B$ 
particles. Driving an $A$ to $B$ transition via radiofrequency
spectroscopy would be a possible way to confirm the 
predicted energies in Fig.~\ref{enetot}. This technique 
has been demonstrated for fermions in recent few-body 
experiments in 1D \cite{wenz2013}.
The results in Fig.~\ref{enetot} show that the energy 
per particle tends to saturate for large systems 
and that this happens faster the 
more imbalanced the system is.
We also see that the balanced
case has an almost linear energy dependence.

\begin{figure}[ht!]
\centering
\includegraphics[scale=0.45,clip=true]{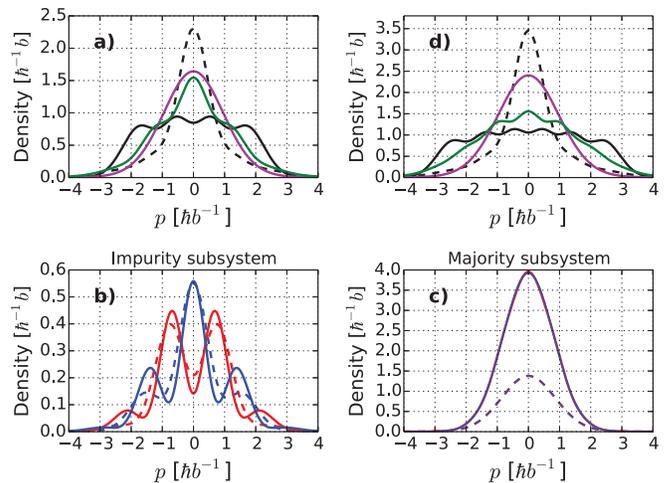}
\caption{Momentum distributions. 
{\bf a}) $N_A=N_B=2$ system. The even parity ground state is shown in purple, while the lowest excited state with 
antiferromagnetic structure is shown in green. For comparison the four-peak black curve represents identical fermions
while the narrow peak dashed black curve shows hard-core bosons.
{\bf b}) Ground state impurity distributions for the even (blue) and odd (red) ground states with $N_B=8$ (solid) and $N_B=3$ 
(dashed). {\bf c}) Ground state majority distributions for $N_B=8$ (solid) and $N_B=3$ (dashed). The even and odd parity 
results coincide for the majority.
{\bf d}) The same as in {\bf a}) for $N_A=N_B=3$. All curves in the 
plot have been obtained for $1/g=0.01$.}
\label{moment}
\end{figure}

The ferro- and antiferromagnetic states can also be 
detected by measuring momentum distributions. 
In Fig.~\ref{moment}{\bf a}) and {\bf d}) we show
the distributions for $1/g=0.01$ with $N_A=N_B=2$ and $N_A=N_B=3$,
respectively. The purple solid line is the even parity ground
state while the solid green line is the excited state with 
antiferromagnetic ordering. The striking difference of the 
two distributions implies that they should be easily 
identifiable in experiments. For comparison, the solid
black line with a multi-peak structure shows the distribution 
for a system of identical fermions. Comparing the solid
green and black curves in Fig.~\ref{moment}{\bf a}) and 
{\bf d}) clearly demonstrates that, in spite of the fact
that these two states have equal energy, the correlations
are very different. In Fig.~\ref{moment}{\bf a}) the dashed
black line corresponds to the Tonks-Girardeau hard-core
boson state \cite{girardeau1960}, which is also seen to be 
very different from the states discussed here. For 
imbalanced systems, we find that measuring the impurity
momentum distribution, Fig.~\ref{moment}{\bf b}), 
yields information about the parity of the state. On the 
other hand, the
majority distributions
in Fig.~\ref{moment}{\bf c}) are identical in the two
opposite parity ground states. 
A characteristic feature
seen in Fig.~\ref{moment}{\bf b}) is the development
of oscillatory structure as the number of majority particles
increases and pushes the impurity further out in the 
trap, see also Fig.~\ref{imdens}{\bf b}).

The separation of components in the ground state is intrinsic
to both balanced and imbalanced mixtures, as is the
presence of other spatial configurations in specific excited states. 
The effect is not connected to the harmonic confinement considered
here and should be seen in an arbitrary confining geometry.
A simple physical picture can be given in terms of domain walls,
i.e. points at which the two components interface. The system 
tends to minimize the number of domain walls and this principle 
can be used to understand the ferromagnetic ground state and 
predict the ordering in energy of other configurations. 
In the paradigmatic two-component (spin 1/2) Fermi system, the 
ground state is never purely ferro- or antiferromagnetic for 
strong interactions \cite{volosniev2013}, and Bose mixtures 
therefore provide a unique set of quantum ground states for
exploring and exploiting magnetic behaviour. 
The description of these 
systems clearly goes beyond the famous Bose-Fermi
mappings \cite{girardeau1960,girar2004} and we provide
not only numerical but also new analytical tools to fill this gap. 
Importantly, we demonstrate 
that the crossover from few- to many-body physics can 
be studied already at the level of ten particles.

\section{Acknowledgements}
This work was funded by the 
Danish Council for Independent Research DFF Natural Sciences and the 
DFF Sapere Aude program and
the European Research Council under the European Community's 
Seventh Framework Programme - ERC grant agreement no. 240603.

\appendix
\section{Numerical method}
We solve numerically the many-body Schr{\"o}dinger equation 
by exact diagonalization with the full Hamiltonian projected 
onto a finite basis constructed from harmonic oscillator 
single-particle states. Each many-body basis state is written 
as a product of symmetrized states of $N_A$ and $N_B$ particles. 
The model space truncation is defined by an upper limit of the total energy. 

Instead of the bare zero-range interaction in \eqref{interaction}, 
we consider an effective two-body interaction in order to speed up 
the convergence of the eigenstates with respect to the size of the 
many-body basis. The effective potential is constructed in a 
truncated two-body space, and is designed such that its solutions 
correspond to exact two-body solutions given by the Busch 
formula~\cite{busch1998}. As explained in detail in 
Refs.~\onlinecite{rotureau2013,lindgren2014}, this is achieved 
using a unitary transformation that involves the lowest 
eigensolutions given by the Busch formula. By construction, 
this unitary transformation approach will reproduce exact 
bare Hamiltonian results for the many-body system (both energy 
spectrum and wave functions) in the limit of infinite model space. 

The excellent convergence property of this effective-interaction 
approach was demonstrated in Ref.~\cite{lindgren2014} and is key 
to the quality of our numerical results and to our conclusions. 
In the construction of the effective interactions we benefit 
from having access to the exact two-body solutions for short-range 
interactions in harmonic traps. However, we stress that using 
numerical two-body solutions this approach can be generalized 
to study many-body systems in higher dimensions with finite-range 
interactions and in any trapping potential.

\section{Analytics for balanced systems}
Here we outline the calculational procedures required to obtain the exact solutions for the $N_A=N_B=2$ four-body
system in the $1/g\to 0$ limit. The method can in principle be extended to larger systems, but it 
becomes increasingly difficult. In the next subsection we provide an alternative method that works well
for larger systems in the imbalanced case.  

Denote the $A$ coordinates by $x_1,x_2$ and the $B$ coordinates by $y_1,y_2$, see Fig.~\ref{2+2}{\bf b}). 
The Hamiltonian is
\begin{align}
H=&\sum_{i=1}^{2}\left[\frac{p_{x_i}^{2}}{2m}+\frac{1}{2}m\omega^2x_{i}^{2}
+\frac{p_{y_i}^{2}}{2m}+\frac{1}{2}m\omega^2y_{i}^{2}\right]&\nonumber\\
&+g\delta(x_1-y_1)
+g\delta(x_1-y_2)+g\delta(x_2-y_1)+g\delta(x_2-y_2),&
\end{align} 
with $g$ the $AB$ interacting coupling constant and we assue that the $AA$ and $BB$ interactions vanish. 
We now perform an orthogonal coordinate
transformation
\begin{align}
\left[\begin{matrix}x \\ y \\ z \\ R\end{matrix}\right]=
\left[\begin{matrix}
\frac{1}{\sqrt{2}}&-\frac{1}{\sqrt{2}}&0&0\\
0&0&-\frac{1}{\sqrt{2}}&\frac{1}{\sqrt{2}}\\
\frac{1}{2}&\frac{1}{2}&-\frac{1}{2}&-\frac{1}{2}\\
\frac{1}{2}&\frac{1}{2}&\frac{1}{2}&\frac{1}{2}
\end{matrix}\right]
\left[\begin{matrix}x_1 \\ x_2 \\ y_1 \\ y_2\end{matrix}\right],
\end{align}
where $x,y,z$ are as shown at the bottom of in Fig.~\ref{2+2}{\bf b}) while $R$ denotes the center-of-mass coordinate.
The quadratic kinetic and harmonic oscillator terms in $H$ retain their form under
this transformation and one may immediately separate the center-of-mass, $R$, which can 
be ignored from now on. For the remaining three coordinates we switch to the 
usual spherical coordinate system, i.e. $\bm r=(x,y,z)$, $r=\sqrt{x^2+y^2+z^2}$, $\tan(\phi)=y/x$, and
$\cos(\theta)=z/r$. Carrying out these transformation we arrive at the relative 
Hamiltonian
\begin{align}\label{hrel}
H_\textrm{rel}=-\frac{\hbar^2\bm \nabla^2}{2m}+\frac{1}{2}m\omega^2\bm r^2+\frac{g}{r}\sum\delta(\cos\theta\pm\sin\theta\sin(\pi/4\pm\phi)),
\end{align}
where the sum in the interaction term runs over the four combinations of signs in the argument of the delta function.
The first two terms constitute a 3D harmonic oscillator with the well-known regular solution
$\Psi(r,\theta,\phi)=N_{nl}r^l e^{-r^2/2b^2}L^{l+1/2}_{(n-l)/2}(r^2/b^2)Y_{lm}(\theta,\phi)$, where $b=\sqrt{\hbar/m\omega}$ and
$L_{a}^{b}(x)$ is the generalized Laguerre polynomial.

When $g=0$, the angular functions, $Y_{lm}(\theta,\phi)$, are the usual spherical harmonic functions with $l$ the 
total and $m$ the projection angular momentum quantum number. However, 
in the limit $1/g\to 0$ we have to enforce non-trivial boundary conditions whenever $A$ and $B$ particles
overlap. Let us focus on the region $z>0$ by restricting to $0\leq\theta\leq\pi/2$ (solutions for $z<0$ may be obtained by symmetry arguments
or by considering instead $\pi/2\leq\theta\leq\pi$). 
The arguments of the Dirac delta-functions in Eq.~\eqref{hrel} vanish when 
\begin{align}
1\pm \tan\theta\sin(\pi/4\pm\phi)=0.
\end{align}
If we define $\tilde\phi=\phi+\pi/4$, we have $1\pm\tan\theta\sin\tilde\phi=0$ and $1\pm\tan\theta\cos\tilde\phi=0$. 
The regions defined by these conditions are illustrated in Fig.~\ref{2+2}{\bf b}). The solid red planes 
show exactly where the arguments of the interaction Dirac delta-functions have to vanish. 

We now make the simple transformation $u=\tan\theta\sin\tilde\phi$ and $v=\tan\theta\cos\tilde\phi$. In 
these new $u,v$ variables, the boundaries are simply $1\pm u=0$ and $1\pm v=0$, i.e. the function must vanish on 
the boundary of a square. Finally, one must transform the angular part of the Laplacian into
the new variables which yields
\begin{align}
&\nabla_{\theta,\phi}\to\nabla_{u,v}=\left(1+u^2+v^2\right)\times&\nonumber\\
&\left((1+u^2)\frac{\partial^2}{\partial u^2}+(1+v^2)\frac{\partial^2}{\partial v^2}+2uv\frac{\partial^2}{\partial u\partial v}
+2u\frac{\partial}{\partial u}+2v\frac{\partial}{\partial v}\right).&
\end{align}
By the procedure outlined above we have transformed the problem of solving a harmonic oscillator problem in a non-trivial 
geometry, i.e. the red region in Fig.~\ref{2+2}{\bf b}), into solving a very simple boundary value problem 
\begin{align}\label{laplace}
\nabla_{u,v}f(u,v)=\lambda(\lambda+1)f(u,v)
\end{align}
with $f(u,v)=0$ for $u,v=\pm 1$. We write the eigenvalue in this way so it matches the usual 3D angular eigenvalue $l(l+1)$.
The equation for $f(u,v)$ may be straighforwardly solved by using a two-dimensional Fourier expansion of the wave function. 
This will produce some spurious solutions as we must also impose bosonic symmetry among the two $A$ and two $B$ particles 
separately. This translates to the requirement that the solution be symmetric when reflected across the two diagonals.
Notice that for each eigenvalue of this problem, $\lambda$, we obtain a whole class of solutions with energies $E/\hbar\omega=2n+\lambda+3/2$
as we may add radial excitations. 

The low-lying solutions of Eq.~\eqref{laplace} are given in Tab.~\ref{datatable}. State number
1, 4, 6, 11, 13, and 15 have the required bosonic symmetries for the balanced system. A number of doubly degenerate
states in the spectrum may be used to construct eigenfunctions for a four-body system with $N_A=3$ and $N_B=1$ (or vice versa).
In this case the wave function must vanish on the diagonal of the $(u,v)$ square domain which is achievable by taking
proper linear combinations. The states marked 'fermions' in Tab.~\ref{datatable} are antisymmetric across the two diagonals
in the $(u,v)$ square and provide allowed states for all four-body two-component Fermi systems, i.e. 2+2, 3+1 or four identical 
fermions. The eigenenergies of the fermionic states have the exact values 7.5, 10.5 and 11.5. Our results differ by
$4\cdot 10^{-5}$ which attests to the accuracy of our method. All states in Tab.~\ref{datatable} have been obtained using 
a modest 400 Fourier basis states. Notice that even though the lowest perfectly antiferromagnetic state for $N_A=N_B=2$ bosons
is at the same energy as the fermionic state number 5 in Tab.~\ref{datatable}, they are not related since states with 
the configurations $ABAB$ or $BABA$ solve a different boundary value problem (corresponding to the yellow regions in Fig.~\ref{2+2}{\bf b})).

\begin{table}
\begin{tabular}{ccc}
State  &  $\lambda+1.5$ &  System  \\
\hline
{\bf 1}	&{\bf 3.88989} & {\bf  2+2 bosons}\\
2	&5.64323 & 3+1 bosons\\
3	&5.64323 & 3+1 bosons\\
{\bf 4}	&{\bf 7.07740} & {\bf 2+2 bosons}\\
5	&7.50002 &  fermions\\
{\bf 6}	&{\bf 7.60232} & {\bf 2+2 bosons}\\
7	&8.76318 &  3+1 bosons\\ 
8	&8.76318 &  3+1 bosons\\
9	&9.52342 &  3+1 bosons\\
10	&9.52342 &  3+1 bosons\\
{\bf 11}	&{\bf 10.21574} & {\bf 2+2 bosons}\\
12	&10.50004 &  fermions\\
{\bf 13}	&{\bf 10.73383} & {\bf 2+2 bosons}\\
14	&11.50004 &  fermions\\
{\bf 15}	&{\bf 11.51706} & {\bf 2+2 bosons}\\
16	&11.85409  &  3+1 bosons\\
17	&11.85409  &  3+1 bosons\\
\end{tabular}
\caption{Low-lying spectrum of Eq.~\eqref{laplace}. The last column denotes the 
specific system for which the particular solution is an allowed eigenstate using the 
notation $N_A+N_B$ for bosons. See the text for details.}
\label{datatable}
\end{table}

The energies obtained using this (semi)-analytical approach for the $N_A=N_B=2$ system are given in Fig.~\ref{2+2}{\bf a}) as triangles at $1/g=0$.
The two lowest triangles correspond to the angular ground state (lowest $\lambda$ value) with $n=0$ and $n=1$. The
two upper triangles are the first and second excited angular solutions both with $n=0$. All four solutions have
the spatial structure $AABB\pm BBAA$. The blue dots in Fig.~\ref{bdens}{\bf a}) show the ground state density 
obtained by the transformation method discussed here.
The rest of the spectrum at $1/g\to 0$ can be obtained by solving the boundary value 
problem in the green ($ABBA\pm BAAB$) and yellow areas ($ABAB\pm BABA$). In the latter case
a fermionized (totally antisymmetric) wave function is a solution.
Our main interest here is to understand the 
ground state so we leave the remaining states and regions for future investigations.

\section{Analytics for imbalanced systems}
The analytics provided here is applied for the Bose polaron, $N_A=1$ and $N_B$ arbitrary, but can be extended to 
other systems. The Hamiltonian can be written
\begin{align}
H=h_0(x_1)+\sum_{i=1}^{N_B}h_0(y_i)+g\sum_{i=1}^{N_B}\delta(x_1-y_i), 
\end{align}
where $h_0(z)=p_{z}^{2}/2m+m\omega^2z^2/2$ is a 1D harmonic oscillator. The $x_1$ coordinate denotes the single $A$ particle, 
the 'impurity', while $y_i$ denotes the coordinates of the majority $B$ particles. 
We introduce an adiabatic decomposition of the total wave function of the form
\begin{align}
\Psi(x_1,y_1,\ldots,y_{N_B})=\sum_j \phi_j(x_1)\Phi_j(y_1,\ldots,y_{N_B}|x_1),
\end{align}
where $\Phi_j$ is a normalized eigenstate of the eigenproblem $\sum_{i=1}^{N_B}h_0(y_i)\Phi_j=E_j(x_1)\Phi_j$
which depends parametrically on $x_1$. This expansion can be related to the Born-Oppenheimer approximation
in which case one may consider $x_1$ the 'slow' variable (typically the nuclear coordinate in molecular physics). 
In the limit of interest $1/g\to 0$, we impose the condition that the total wave function vanishes
for $y_i=x_1$, $i=1,\ldots,N_B$. This implies that $\Phi_j=0$ whenever $y_i=x_1$. 
Since there are no intra-species interactions among the $B$ particles, we can write 
\begin{align}
\Phi_j(y_1,\ldots,y_{N_B}|x_1)=S\prod_{i=1}^{N_B}f_{k_i}(y_i|x_1),
\end{align}
where $S$ denotes the symmetrization operator and $f_{k_i}(y_i|x_1)$ is the $k$th normalized eigenstate of $h_0(y_i)$ which 
satisfies the condition $f_{k_i}(y_i=x_1)=0$. The index $j$ on $\Phi_j$ denotes the many different ways to 
distribute the $N_B$ particles among the eigenstates of $h_0(y_i)$ with the appropriate boundary condition.
The Schr{\"o}dinger equation for $\phi_j(x_1)$ can now be written
\begin{align}
\left[h_0(x_1)+E_i(x_1)\right]\phi_i=\sum_j \left(Q_{ij}(x_1)\phi_j+P_{ij}(x_1)\frac{\partial \phi_j}{\partial x_1}\right),
\end{align}
where 
\begin{align}
&P_{ij}(x_1)=\langle \Phi_i|\frac{\partial}{\partial x_1}|\Phi_j\rangle_{y}&\\
&Q_{ij}(x_1)=\frac{1}{2}\langle\Phi_i|\frac{\partial^2}{\partial x_{1}^{2}}|\Phi_j\rangle_{y}.&
\end{align}
The subscript $y$ on the brackets denote integration over all $y_1,\ldots,y_{N_B}$. Note that
$P_{ii}=0$ and $Q_{ii}<0$ \cite{nielsen2001}. 

As we are interested in the ground state, we assume that all the $B$ particles are in the same
state, $f_0(y_i|x_1)$, that we specify below. Since the nearest excited states are obtained by 
promoting one of the $B$ particles into a
single-particle excited orbital, one can show that $P_{ij}$ and $Q_{ij}$ scales with $\sqrt{N_B}$ while
$Q_{ii}$ scales with $N_B$. For large $N_B$ we can thus neglect all but the $Q_{ii}$ terms.
Furthermore, in the ground state we expect to find all the $B$ particles on one side of 
the impurity. If we assume that all $B$ particles are to the left of the impurity, we can
write the single-particle wave function, $f_0(x|z)$ for $x\leq z$, as 
\begin{align}
f_0(x|z)=A(z)e^{-x^2/2}U(1/4-\epsilon(z)/2,1/2,x^2),
\end{align}
for $z<0$ or $z\geq 0$ and $x\leq 0$, while for $z\geq 0$ and $x>0$ we write
\begin{align}
&f_0(x|z)=-A(z)e^{-x^2/2}U(1/4-\epsilon(z)/2,1/2,x^2)&\nonumber\\
&+2A(z)e^{-x^2/2}\frac{U(1/4-\epsilon(z)/2,1/2,0)}{M(1/4-\epsilon(z)/2,1/2,0)} M(1/4-\epsilon(z)/2,1/2,x^2).&
\end{align}
Here $A(z)$ is a normalization factor, $U$ and $M$ are the Tricomi and Kummer confluent
hypergeometric functions, and we have used $b=\sqrt{\hbar/m\omega}$ as the unit of length.
Here $\epsilon(z)$ is a function that is chosen to satisfy the requirement $f(x=z)=0$.
This is equivalent to finding the ground state solution of $h_0(x)$
for $x\leq z$
with the condition that the wave function must vanish at $z$.

Once we have determined the functions $f(x|z)$ and $\epsilon(z)$, we can compute the adiabatic 
potential for the ground state. We have
\begin{align}
Q_{11}(x_1)=-\frac{1}{2}N_B\langle\left(\frac{\partial f(y|x_1)}{\partial x_1}\right)^2\rangle_{y}.
\end{align}
Furthermore, $E_1(x_1)=N_B\epsilon(x_1)$ by additivity.
The Schr{\"o}dinger
equation for $\phi(x_1)$ is then
\begin{align}
\left(h_0(x_1)+N_B\epsilon(x_1)+N_B\langle\left(\frac{\partial f(y|x_1)}{\partial x_1}\right)^2\rangle_y\right)\phi(x_1)=E_0\phi(x_1).
\end{align}
The energy $E_0$ provides a variational upper bound to the exact energy.

The energies computed via this method for the polaron are shown in Fig.~\ref{enetot} and
agree with the numerical results to within a few percent for the largest particle numbers
in the figure. We expect the agreement to become better for even larger particle numbers. 
Furthermore, since we obtain the full wave function in an analytical form we may also 
compute the densities of both impurity and majority components. In Fig.~\ref{imdens}{\bf a})
we show the impurity densities for $N_B=3$ and $N_B=8$, while Fig.~\ref{imdens}{\bf d})
shows the corresponding majority density. We see a striking agreement between the numerical
results and the analytically tractable model presented here. The model presented here can 
be extended to excited states and also to systems with $N_A>1$.

\end{document}